\documentstyle[pre,aps,amsmath,psfig]{revtex}

\newcommand{\K}{{\mathrm{K}}}
\newcommand{\I}{{\mathrm{I}}}

\begin{document}
\title{Meandering instability of curved step edges on growth of a crystalline cone}
\author{M. Rusanen$^{1,2}$, I. T. Koponen$^{2}$ and T. Ala-Nissila$^{1,3}$}
\address{$^{1}$Helsinki Institute of Physics and Laborarory of Physics, Helsinki\\
University of Technology, P.O. Box 1100, FIN-02015 HUT, Espoo, Finland.\\
$^{2}$Department of Physics, P.O. Box 64, FIN-00014 University of Helsinki, Finland\\
$^{3}$Department of Physics, Brown University, Providence R.I. 02912--1843, U.S.A.\\
}

\twocolumn

\maketitle

\begin{abstract}
We study the meandering instability during growth of an isolated
nanostructure, a crystalline cone, consisting of concentric circular
steps. The onset of the instability is studied
analytically within the framework of the standard Burton-Cabrera-Frank
model, which is applied to describe step flow growth in circular geometry.
We derive the correction to the most unstable wavelength and show
that in general it depends on the curvature in a complicated way. Only in the
asymptotic limit where the curvature approaches zero the results are shown to reduce
to the rectangular case. The results obtained here are of importance in estimating
growth regimes for stable nanostructures against step meandering.
\end{abstract}


\section{Introduction}

Film growth by Molecular Beam Epitaxy (MBE) is essentially based on the possibility to
control growth on a submonolayer level. Usually one aims at surfaces
as smooth as possible with atomistically sharp structures consisting
of step edges or nanoscale islands. Growth of these structures in MBE is affected
not only by stochastic but also by deterministic instabilities such as
step meandering and bunching. Since the advent of modern film growth techniques these
surface instabilities have been of theoretical and practical
interest~\cite{jeong00,politi00}.

To obtain nanostructures in MBE with desired quality one of the fundamental
concerns is the stability of step edges. 
Instabilities lead to nonuniform layers so it is advantageous
to suppress them during growth. In step flow growth the basic mechanism
and properties of the meandering instability were treated by Bales and Zangwill
(BZ)~\cite{bz90} on the basis of the classic Burton-Cabrera-Frank (BCF)
model~\cite{bcf51}.
The case of circular nanostructures has so far received much less
attention although it is an example for a layered nanostructure~\cite{fin00}. Decay of
mesoscopic circular stepped structures can be described in some special cases of
stable step flow with good accuracy using the BCF model~\cite{jeong00,fin00}. Also,
recently the decay and bunching of a circular crystalline cone 
in the stable growth regime
has been studied in detail using a similar approach~\cite{kandel99}. However, to our
knowledge there has been no study of the 
meandering instability in the spirit of BZ
during growth of circular nanostructures even though its 
importance was noted already
a decade ago~\cite{bz90}. It is thus of interest to know under which conditions
circular step edges are stable against meandering. In this report we present our
study of the morphological instability of curved step edges. Our viewpoint is based on
the BCF model and we generalize the BZ results to the case of
a circular geometry. We perform a
linear stability analysis and calculate the corrections to the results 
for the straight steps. As expected,
our results reduce to the BZ case in the limit of an infinite step radius.

\section{Step flow growth in circular geometry}

We study a model of circular steps which are placed
concentrically on top of each other. The steps can absorb and emit atoms which
diffuse on terraces between the steps. The terrace $j$ is bounded by the steps at
$r_{j}$ and $r_{j+1}$. Assuming the flux of adatoms onto and evaporation from the
terraces, the adatom concentration $c_{j}$ on the terrace $j$ obeys the well
known form of the BCF equation~\cite{bz90,bcf51}
\begin{equation}
\label{eq:bcf}
\frac{\partial c_{j}}{\partial t} = D_{s}\nabla^{2}c_{j} - c_{j}/\tau + F,
\end{equation}
where $D_{s}$ is the diffusion coefficient of an adatom on a flat terrace,
$\tau$ is the time scale for evaporation, and $F$ is the deposition flux.
Assuming that the adatom concentration relaxes much faster than the step edge
moves we can assume that the terrace is in a quasi-stationary state
corresponding to $\partial c_{j}/\partial t = 0$. Mass transport through the bulk
of the material is ignored.

The model is now fully specified with the choice of the boundary
conditions and the requirement of the mass conservation at the step edges.
The mass conservation implies that the edge velocity is given by~\cite{bz90}
\begin{equation}
\label{eq:mass}
V_{j} = V_{j+} + V_{j-} =
\Omega D_{s}(\frac{\partial c_{j}}{\partial r}\lvert_{R_{j}}
   - \frac{\partial c_{j-1}}{\partial r}\lvert_{R_{j}}),
\end{equation}
where $\Omega$ is the atomic area, $V_{j-},V_{j+}$ are the
contributions to the step edge velocity due to surface
currents from the upper and lower terrace, respectively, and $R_j$ is the
radius of the $j$th step edge. We assume that the
velocities $V_{j \mp}$ are related to the deviations of the adatom
concentration from the equilibrium value~\cite{chernov}
\begin{equation}
\label{eq:dev}
V_{j \pm} = \Omega k_{\pm}[c_{j} - c_{j}^{eq}],
\end{equation}
where $c_{j}^{eq} = c_{0}^{eq}\exp [\tilde{\Gamma} \kappa(R_j)]$ is the equilibrium
adatom concentration at the edge, $\tilde{\Gamma} = \Omega \gamma /(k_B T)$,
$\gamma$ is the free energy/(unit length), $c_{0}^{eq}$ is the equilibrium adatom
concentration at the straight step, $\kappa(R_j)$ is the local curvature of the
step in the circular geometry (see e.g. Eq. (7) in Ref.~\cite{miranda98}), and
$k_{-},k_{+}$ are the attachment coefficients associated with the upper
and lower terraces, respectively. From 
Eqs.~(\ref{eq:mass}) and (\ref{eq:dev}) we obtain the mixed
boundary conditions at the step edge
\begin{eqnarray}
\label{eq:mixed}
 D_{s}\frac{\partial c_{j}}{\partial r}\lvert_{R_{j}} & = &
           k_{+}[c_{j}\lvert_{R_{j}} - c_{j}^{eq}], \nonumber \\
-D_{s}\frac{\partial c_{j}}{\partial r}\lvert_{R_{j+1}} & = &
           k_{-}[c_{j}\lvert_{R_{j+1}} - c_{j+1}^{eq}].
\end{eqnarray}

Defining a new field $u_{j} = c_{i}-\tau F$ Eq.~(\ref{eq:bcf}) becomes
the Helmholtz equation in the stationary limit~\cite{bz90,bcf51}:
\begin{equation}
\label{eq:scaled}
\nabla^{2} u_{j} - \frac{1}{x_{s}^{2}} u_{j} = 0,
\end{equation}
where $x_{s} = \sqrt{D_{s}\tau }$ is the diffusion length.
The solution of Eq.~(\ref{eq:scaled}) for the perfectly circular step is given by
$u_{j}^0(r) = a_{j}^0 \I_{0}(r/x_{s}) + b_{j}^0 \K_{0}(r/x_{s})$, 
where $\I_{0}(x)$
and $\K_{0}(x)$ are the zeroth order modified Bessel functions~\cite{absteg60}
and $a_{j}^0$ and $b_{j}^0$ are coefficients determined by the boundary conditions.
In the linear stability analysis a small perturbation is added to the step edge
and the equations are solved to first order in the perturbation amplitude.
We set
$\tilde{r}_j(\theta ) = R_j + \epsilon \exp [in\theta +\omega t]+{\mathrm{c.c.}},$
where $\epsilon$ is a small parameter, $\omega $ a growth rate, $|n| \ge 1$ is
an integer (due to periodicity), and c.c. denotes the complex conjugate.
The solution to Eq.~(\ref{eq:scaled}) in the first order in $\epsilon$ gives
$u_j(r,\theta ) = u_{j}^0(r) + \epsilon [{A}_{j}^n \I_{n}(r/x_{s})
 + B_{j}^n \K_{n}(r/x_{s})]e^{in \theta +\omega t}$, 
where $\K_{n}(x)$ and $\I_{n}(x)$
are the modified Bessel functions of integer order $n$, and the coefficients
$A_{j}^n$ and $B_{j}^n$ are determined by the boundary conditions.
When the solution is found, the growth rate $\omega$ can be deduced using
Eq.~(\ref{eq:mass}) and $V_j = V_{j}^0 + \omega \tilde{r}_j(\theta)$. If $\omega > 0$
the step edge is linearly unstable.

\section{Morphological instability of a circular step}

Qualitatively, the growth of stepped structures is unstable against step meandering
when the flux of adatoms from
the upper terrace is reduced, {\it e.g.} 
due to the Ehrlich-Schwoebel barrier. This is
basically the origin of the morphological instability on vicinal surfaces (the BZ
case)~\cite{bz90}. However, in the case of a circular step, the stabilizing effect
of the step curvature is expected to be more pronounced than in the rectangular
geometry. Therefore, we expect the possible instability to be weaker than in the
rectangular case since the line tension tends to smoothen the steps. Here
we consider the cases $k_{+} \rightarrow \infty$, $k_{-} = 0$ (one-sided model),
and $k_{+} \neq k_{-}$ non-zero and finite (asymmetric model).

\bigskip
\subsection{One-sided model}

In the one-sided model $k_{+} \rightarrow \infty $ corresponds to instantaneous
attachment from the lower terrace and $k_{-} = 0$ implies an infinite
Ehrlich-Schwoebel barrier. In this limit the velocity of the step with radius
$R$ is given by $V = V_{+} = D\Omega (\partial u/\partial r)$ and the stability
function becomes
\begin{equation}
\label{eq:omega_one}
\frac{\omega (n)}{\Omega \Delta F} = (\frac{\xi_{s}}{\rho}-1)
  \frac{b_{n}^{1} + b_{n}^{2} + b_{n}^{3}}{a_{n}}
       + c_{n}\frac{\xi_{s}}{\rho^{2}}(1-n^{2}),
\end{equation}
where $\xi_{s} = \tilde{\Gamma} /(x_{s}\tau \Delta F)$ is the capillary length,
$\Delta F=F-c_{0}^{eq}\tau$, and $\rho = R/x_{s}$. The coefficients are given by
$a_{n} =  [\widehat{\I_{n}^{\prime}}\K_{n} - \I_{n}\widehat{\K_{n}^{\prime}}]
          [\widehat{\I_{1}}\K_{0} + \I_{0}\widehat{\K_{1}}]$,
$b_{n}^{1} = [\widehat{\I_{n}^{\prime }}\K_{n} - \I_{n}\widehat{\K_{n}^{\prime }}]
            [\I_{1}^{\prime }\widehat{\K_{1}} - \widehat{\I_{1}}\K_{1}^{\prime }]$,
$b_{n}^{2} = [\I_{n}^{\prime }\widehat{\K_{n}^{\prime }}
              - \widehat{\I_{n}^{\prime }}\K_{n}^{\prime }]
            [\I_{1}\widehat{\K_{1}} - \widehat{\I_{1}}{K}_{1}]$,
$b_{n}^{3} = [\I_{n}\K_{n}^{\prime} - \I_{n}^{\prime }\K_{n}]
            [\widehat{\I_{1}^{\prime }}\widehat{\K_{1}}
               - \widehat{\I_{1}}\widehat{\K_{1}^{\prime }}],$ and
$c_{n} = [\widehat{\I_{n}^{\prime }}\K_{n} - \I_{n}\widehat{\K_{n}^{\prime}}]
        [\widehat{\I_{n}^{\prime }}\K_{n} - \I_{n}\widehat{\K_{n}^{\prime }}]$,
where $\I_{n} \equiv \I_{n}(\rho)$, $\widehat{\I_{n}} \equiv \I_{n}(\rho+l/x_{s})$,
$l$ is the terrace width, and
the prime indicates the derivative with respect to the scaled variable $\rho$.
By using the asymptotic formulae of the modified Bessel functions for
$n,R$ large with $n/R\equiv q = const.$ \cite{absteg60} the growth
rate (\ref{eq:omega_one}) can be shown to reduce to the result of Bales and
Zangwill~\cite{bz90} in the limit $R \rightarrow \infty$.

The stability function is plotted in Fig.~1 and it approaches a limiting form
when the radius of the step increases (curvature
decreases). As can be seen from the figure, there exists a limiting value $R_c$
for the radius such that steps with $R < R_c$ are always stable. For radii $R > R_c$
there exists a critical value of the wavevector $q_c$ such that for $q > q_c$ the
edge is stable. The critical wavevector depends on curvature and with
increasing curvature $q_{c}$ decreases, {\it i.e.} the critical wavelength where the
instability sets in is shifted to larger wavelengths.

In the limit of large $n,R$ we obtain the correction term to the results of
Bales and Zangwill of the order of $1/R$ as
\begin{equation}
\frac{\omega (n)}{\Omega \Delta F} = \omega_{BZ}(q) + \frac{x_s}{2R}\Sigma_q,
\end{equation}
where $\omega_{BZ}$ is the rectangular result (Eq.~(13) in Ref.~\cite{bz90}),
\begin{eqnarray}
\Sigma_q & = &  [x_s l q^2 \Lambda_{q}^{-2} 
      + \Lambda_{q}^{-1}\tanh(\Lambda_q\tilde{l})(l^2 q^2 + \Lambda_{q}^{-2})
   + \tanh(\tilde{l}) \nonumber  \\
  && - 2\xi_s ]
      \times\frac{1}{\cosh(\Lambda_q\tilde{l})\cosh(\tilde{l})} \nonumber \\
&& -\tanh(\tilde{l})(2+l^2 q^2) + 
(l^2 q^2 + \Lambda_{q}^{-2})\tanh^2(\Lambda_q\tilde{l})\tanh(\tilde{l}) \nonumber \\
&&+ 2\xi_s 
 + \xi_s x_s^2 l^2 q^4 - 
\xi_s x_s q^2 (l^2 q^2 + \Lambda_{q}^{-2})\tanh^2(\Lambda_q\tilde{l}), \nonumber
\end{eqnarray}
$\Lambda_q = \sqrt{1+(x_s q)^2}$, and $\tilde{l}=l/x_s$. Using the equation above
we can determine analytically the corrections to

\psfig{file=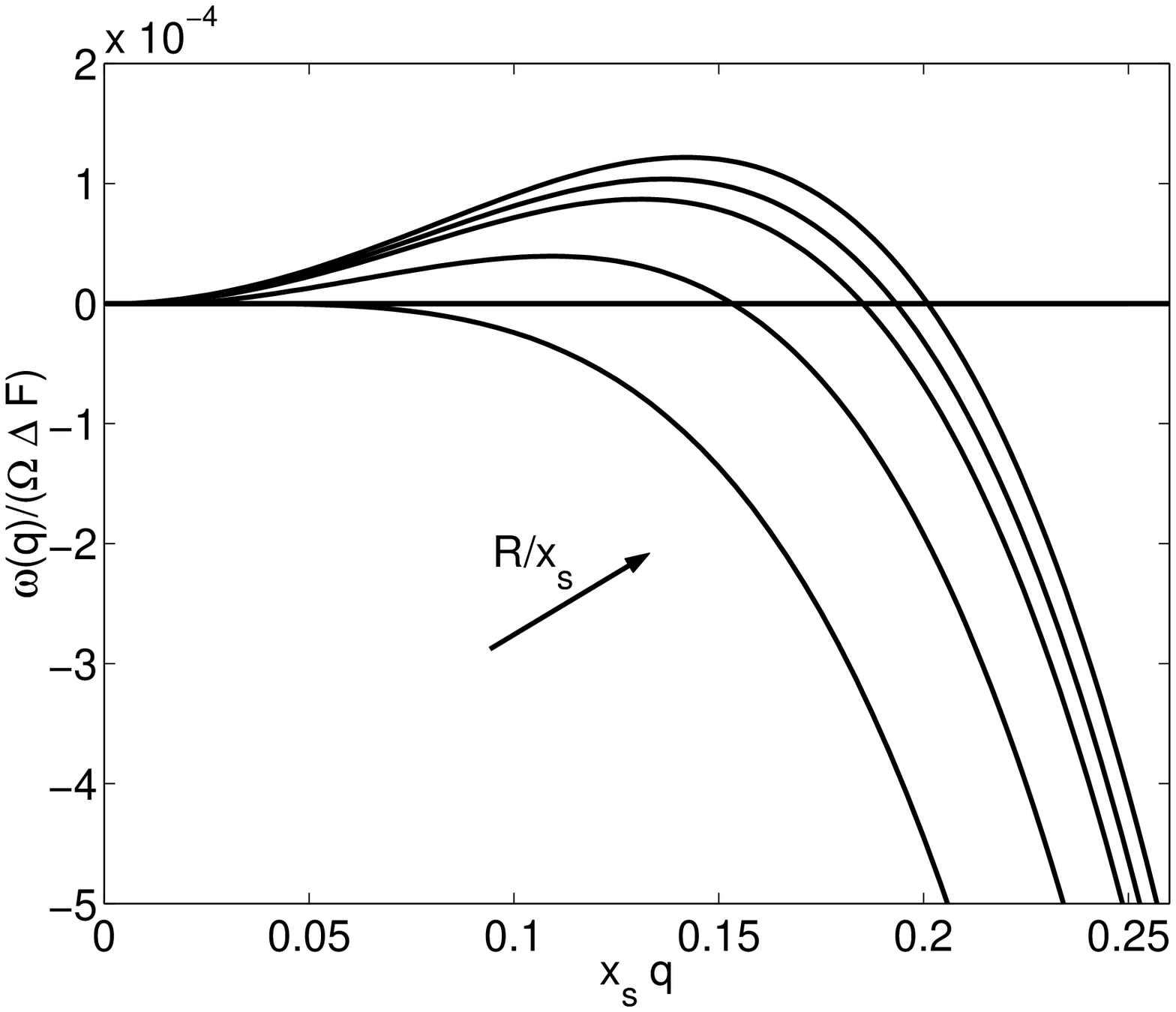,width=8.2cm}
\bigskip
\noindent {\bf Fig. 1:} The growth rate $\omega(q)$ as a function of the 
wavevector $q=n/R$ with a moderate value
of the terrace width $l$ in the case of one-sided model. The stable regime
corresponds to $\omega < 0$. The radius of the
step $R$ increases in the direction of the arrow.
The curves indicate that structures smaller than a critical size
$R_c$ are stable against step meandering.

\bigskip
\bigskip

\noindent the critical wavevector defined as $\omega(q_c) = 0$. The results
are for $l \gg x_s$:
\begin{equation}
\label{eq:appr1}
x_{s}q_{c}\approx
    \begin{cases}
\sqrt{\frac{1}{\xi_s} + \frac{x_s}{2R}(\frac{1}{\xi_s} - 2) }   &, x_s q_c \gg 1; \\
\sqrt{\frac{4}{3}(1-2\xi_s)-\frac{2x_s}{R} } &, x_s q_c \ll 1,
    \end{cases}
\end{equation}
and for $l\ll x_{s}$ (and $l^2 q_c^2 \ll 1$):
\begin{equation}
\label{eq:appr2}
x_ s q_c \approx \sqrt{\frac{lx_s}{2\xi_s}-1 - \frac{l}{R})}.
\end{equation}
Omitting the $1/R$ terms we obtain the BZ results for the rectangular
geometry~\cite{bz90}.
In Fig.~2 the curves for the critical wavevector $q_c$ against the capillary
length $\xi_s$ are shown for $l \gg x_s$, $x_s q_c \ll 1$. The corrected
result follows the numerically plotted curve whereas the BZ result deviates
considerably. For $\xi_s$ large enough the edge is always stable and $q_c$
aprroaches zero. The inset of Fig.~2 shows the case $l \ll x_s$ which behaves
in a similar way.

\subsection{Asymmetric model}

When the kinetic coefficients at the step edge from the lower and upper
terrace are both finite and non-zero, the velocity of the $j$th step edge is given
by $V_{j} = D_{s}\Omega \lbrack (\partial u_{j}/\partial r) -
             (\partial u_{j-1}/\partial r)]\lvert_{r=\tilde{r}_j(\theta )}$
and the growth rate becomes

\psfig{file=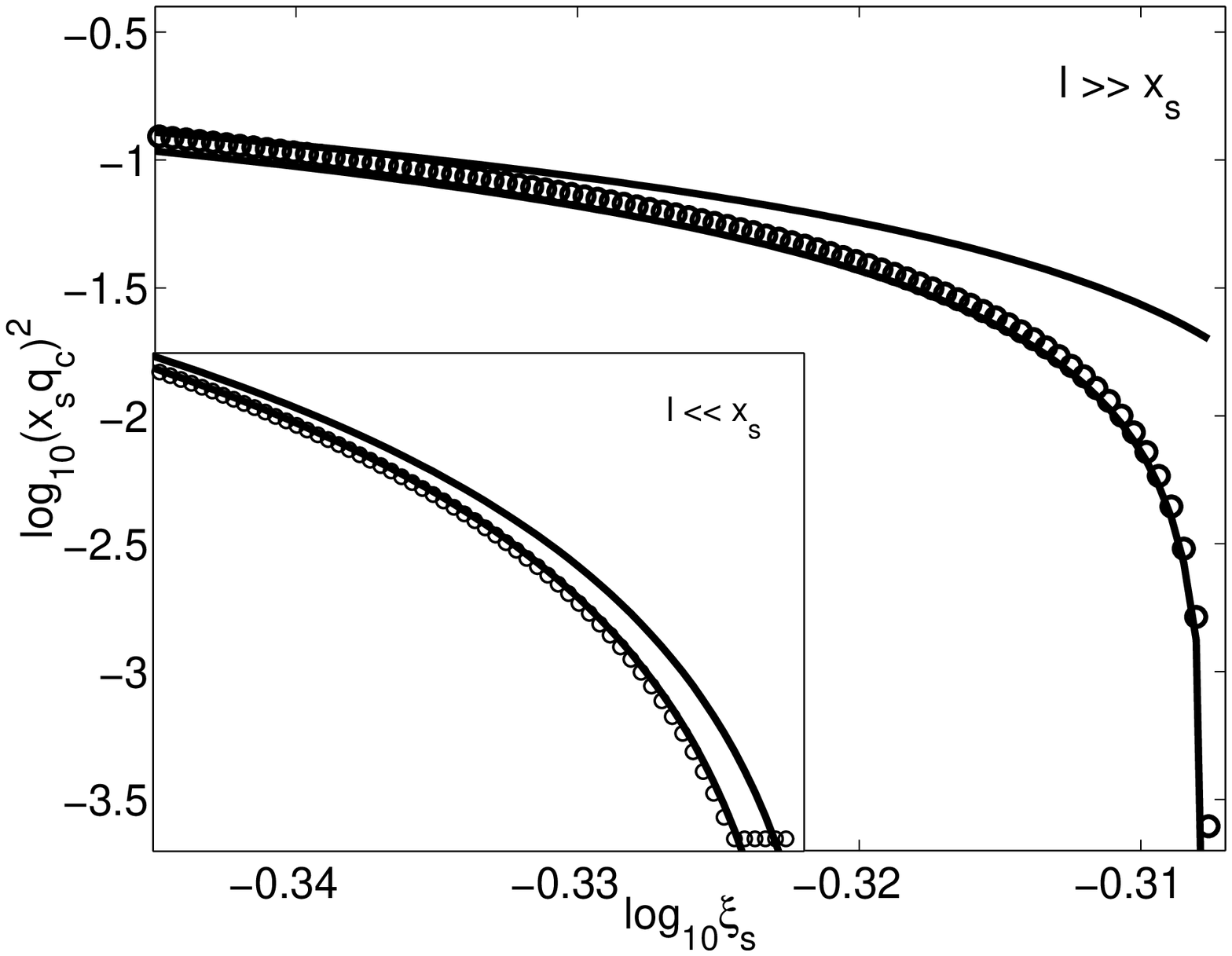,width=8.2cm}
\bigskip
\noindent {\bf Fig. 2:} The critical wavevector $q_c$ as a function
of the capillary length $\xi_s$ in the case of the one-sided model.
The circles are from of Eq.~(\ref{eq:omega_one}),
the upper solid line is the BZ result~\cite{bz90}, and the lower solid line
is Eq.~(\ref{eq:appr1}) for $l \gg x_s$. The inset shows the case $l \ll x_s$,
with the lower solid line given by Eq.~(\ref{eq:appr2}).
The approximations follow the numerically plotted curve, whereas the BZ results
deviate for larger values of $\xi_s$.

\bigskip
\bigskip

\begin{eqnarray}
\frac{\omega (n,j)}{\Omega \Delta F} & = &
    [\frac{\alpha_{j}}{\gamma_{j}} - \frac{\alpha_{j-1}}{\gamma_{j-1}} ]
         \, \I_{0}^{\prime \prime}(\rho_{j})
  + [\frac{\beta_{j}}{\gamma_{j}} - \frac{\beta_{j-1}}{\gamma_{j-1}} ]
         \, \K_{0}^{\prime \prime}(\rho_{j})  \\
&+& [ \frac{{\cal A}_{n}^{j}}{{\cal D}_{n}^{j}}
       - \frac{{\cal A}_{n}^{j-1}}{{\cal D}_{n}^{j-1}} ]
        \, \I_{n}^{\prime }(\rho_{j})
  + [ \frac{{\cal B}_{n}^{j}}{{\cal D}_{n}^{j}}
       - \frac{{\cal B}_{n}^{j-1}}{{\cal D}_{n}^{j-1}} ]
        \, \K_{n}^{\prime }(\rho_{j}), \nonumber 
\end{eqnarray}
where $\alpha_j$, $\beta_j$, and $\gamma_j$
are the coefficients related to unperturbed steps, and
${\cal A}_{n}^j$, ${\cal B}_{n}^j$, and ${\cal D}_{n}^j$
are obtained from the expansion 
linear in $\epsilon$. Define $\tilde{d}_{\pm}=d_{\pm}/x_s=(D_s/k_{\pm})/x_s$,
${\cal K}_{n,\pm }^j = \K_{n}(\rho_{j}) \mp \tilde{d}_{\pm }\K_{n}^{\prime}(\rho_{j})$
and ${\cal I}_{n,\pm }^j = 
\I_{n}(\rho_{j}) \mp \tilde{d}_{\pm }\I_{n}^{\prime}(\rho_{j})$.
Then the coefficients are given by
\begin{eqnarray}
\alpha_{j} & = &  (\xi_s / \rho_{j} - 1){\cal K}_{0,-}^{j+1}
                - (\xi_s / \rho_{j+1} - 1){\cal K}_{0,+}^j, \nonumber \\
\beta_{j} & = & - (\xi_s / \rho_{i}-1){\cal I}_{0,-}^{j+1}
                + (\xi_s / \rho_{i+1} - 1){\cal I}_{0,+}^j,  \nonumber \\
\gamma_{j} & = & {\cal I}_{0,+}^{j}{\cal K}_{0,-}^{j+1}
                   - {\cal I}_{0,-}^{j+1}{\cal K}_{0,+}^{j}, \nonumber \\
{\cal A}_{n}^j & = & \alpha_j [ {\cal I}_{1,-}^{j+1}{\cal K}_{n,+}^j
                               - {\cal I}_{1,+}^j{\cal K}_{n,-}^{j+1} ]
            + \beta_j [ {\cal K}_{1,+}^j{\cal K}_{n,-}^{j+1}
                       - {\cal K}_{1,-}^{j+1}{\cal K}_{n,+}^{j} ] \nonumber \\
&&-\xi_s(n^2-1)[ {\cal K}_{n,+}^{j}/\rho_{j+1}^2 - {\cal K}_{n,-}^{j+1}/\rho_{j}^2 ],
	\nonumber \\
{\cal B}_{n}^j & = & - \alpha_j [ {\cal I}_{1,-}^{j+1}{\cal I}_{n,+}^j
                               - {\cal I}_{1,+}^j{\cal I}_{n,-}^{j+1} ]
            - \beta_j [ {\cal K}_{1,+}^j{\cal I}_{n,-}^{j+1}
                       - {\cal K}_{1,-}^{j+1}{\cal I}_{n,+}^{j} ] \nonumber \\
&&+\xi_s(n^2-1)[ {\cal I}_{n,+}^{j}/\rho_{j+1}^2 - {\cal I}_{n,-}^{j+1}/\rho_{j}^2 ],
	\nonumber \\
{\cal D}_{n}^j & = & - {\cal I}_{n,-}^{j+1}{\cal K}_{n,+}^j
                     + {\cal I}_{n,+}^{j}{\cal K}_{n,-}^{j+1}. \nonumber
\end{eqnarray}

The resulting expression for the growth rate $\omega(q)$
is rather complicated and we have not found any essentially simpler form
which would be accurate enough for all curvatures. However, the essential
qualitatitive features can be extracted from the special cases, and
the
\psfig{file=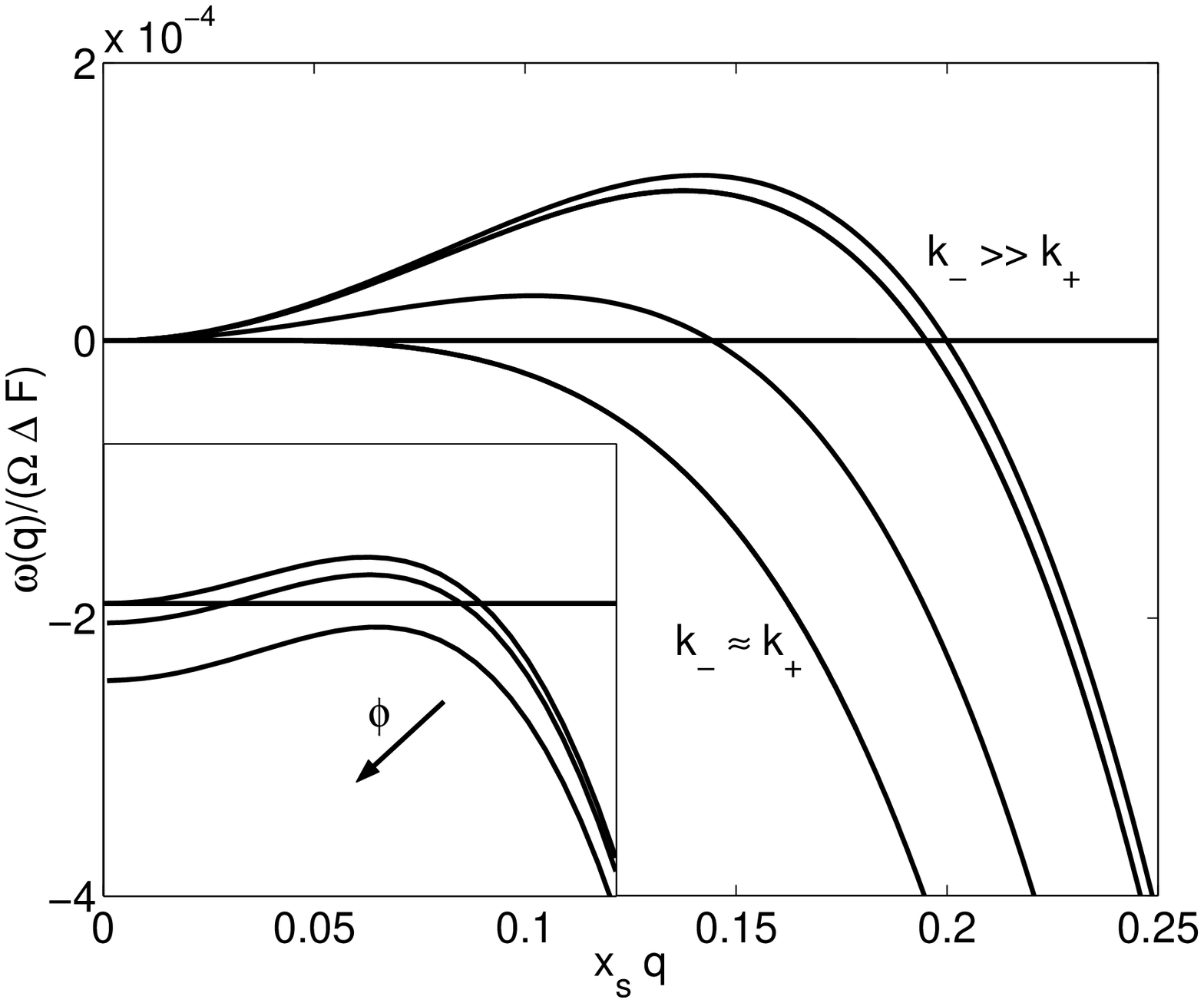,width=8.2cm}
\bigskip
\noindent {\bf Fig. 3:} The instability function $\omega(q)$ as a function of
the wavevector $q$ in the asymmetric case. The attachment coefficients $k_{-}$,
$k_{+}$ approach the same value from top to bottom. For $k_{-} \leq k_{+}$
the steps are always stable.
The inset shows $\omega$ with increasing phase difference $\phi$
between two adjacent steps. Increasing $\phi$ makes steps more stable
against meandering as in the rectangular case~\cite{pimpinelli94}.

\bigskip
\bigskip

\noindent behavior is similar to the one-sided case. The numerically plotted values of
$\omega$ indicate that as $k_{-}$ approaches $k_{+}$ the growth rate $\omega \leq 0$
at all values of $q$. This is shown in Fig.~3.
We can thus conclude that the one-sided model is
the most unstable.

So far, we have only considered the case of in-phase step
growth. The stability of growth depends also on the phase of the
neighboring steps, and the above analysis can be extended to the more general
situation with an arbirary phase between adjacent steps.
However, the general results give only little additional insight and only the
numerical results are shown in the inset of Fig.~3. Increasing the
phase difference between the two adjacent steps makes the steps more stable
as in the case of rectangular geometry~\cite{pimpinelli94}.

\section{Discussion and conclusions}

In this work we have generalized the 
meandering instability to structures made of
concentric islands. We have shown that in the case of circular cones growth
instability due to terrace diffusion can arise as in the case of rectangular geometry.
However, the instability is suppressed by the curvature and structures
with smaller sizes than a critical size are stable. The critical size
depends on the microscopic parameters of the system. We present here also the
asymptotic analytical results for small curvatures. In the
limit $R \rightarrow \infty$ the growth rate approaches the expression found by
Bales and Zangwill~\cite{bz90}. The more general case where attachment from the upper
and lower terraces to the step are both finite and non-zero behaves
qualitatively similarly to the one-sided model.

The results presented here extend the results of the BZ model for rectangular geometry
and they are of interest in modeling {\it e.g.} the behaviour of isolated
crystalline cones. Results given here can be used to check whether
the simplified model which assumes the perfectly circular step edges in
decaying cones is valid. If the meandering instability is
expected to be present, it can play a significant role in the evolution of
nanostructures.
\bigskip

Acknowledgements: We acknowledge the Academy of Finland for financial support,
in part through its Center of Excellence program. M.R.
also thanks Magnus Ehrnrooth foundation for travel support.
\medskip


\end{document}